\documentclass[12pt]{article}
\usepackage{authblk}
\usepackage{natbib}
\usepackage{amsmath, amsfonts,mathrsfs}
\usepackage{color}
\usepackage{rotating}
\usepackage{JASA_manu1}
\usepackage{lscape}
\usepackage{graphicx,multirow}
\usepackage{bm,url}
\usepackage{titling,subfigure}
\usepackage{float}
\usepackage{longtable}

\usepackage{amsmath}
\allowdisplaybreaks[1]

\newtheorem{lemma}{\noindent\mbox{Lemma}}
\newtheorem{theorem}{\noindent\mbox{Theorem}}
\newtheorem{corollary}{\noindent\mbox{Corollary}}
\newtheorem{proposition}{\noindent\mbox{Proposition}}
\newtheorem{condition}{\noindent\mbox{M}}

\def\be{\begin{equation}}
\def\ee{\end{equation}}
\def\bea{\begin{eqnarray}}
\def\eea{\end{eqnarray}}
\def\bd{\begin{displaymath}}
\def\ed{\end{displaymath}}
\def\bda{\begin{eqnarray*}}
\def\eda{\end{eqnarray*}}
\def\bsm{\begin{small}}
\def\esm{\end{small}}

\newcommand{\E}{{\rm E}}

\def\ha1{\hat \beta_1}

\def\bb0{\delta_\beta}

\def\bsc{\begin{scriptsize}}
\def\esc{\end{scriptsize}}

\begin{document}
\title{Max-Type and Sum-Type Procedures for Online Change-Point Detection in the Mean of High-Dimensional Data}
\author{Jun Li}
\affil{Department of Mathematical Sciences, Kent State University, Kent, OH 44242\\
Email: {jli49@kent.edu} }
\date{}

\renewcommand\Affilfont{\itshape\normalsize}
\maketitle

\begin{center}
\textbf{Abstract}
 \end{center}
We propose two procedures to detect a change in the mean of high-dimensional online data. One is based on a max-type U-statistic and another is based on a sum-type U-statistic. 
Theoretical properties of the two procedures are explored in the high dimensional setting. More precisely, we derive their average run lengths (ARLs) when there is no change point, and expected detection delays (EDDs) when there is a change point.   
Accuracy of the theoretical results is confirmed by simulation studies. 
The practical use of the proposed procedures is demonstrated by detecting an abrupt change in PM2.5 concentrations. The current study attempts to extend the results of the CUSUM and Shiryayev-Roberts procedures previously established in the univariate setting.


\setcounter{section}{1} \setcounter{equation}{0}
\section*{\large 1. \bf Introduction}

Let $X_1, X_2, \cdots$ be a sequence of independent observations, and let 
$\tau$ be a time point where the distribution of observations is changed.  An important task is to detect the change point $\tau$ as observations are continuously arriving. In the sequential setting, the change point $\tau$ is identified by the stopping time $T$ of a stopping rule, which decides whether to terminate or continue a monitored process based on the current and past observations. There are two errors to be controlled for a stopping rule. One is false positive, meaning that the stopping rule wrongly produces a stopping time if there is no change ($\tau=\infty$). Another is a detection delay which is the number of additional observations the stopping rule needs in order to detect the change point $\tau$. A desirable stopping rule should maintain the expected detection delay (EDD) as small as possible, subject to the constraint that the expected false positive or the average run length (ARL) is greater than a pre-specified constant.     

Sequential change point detection has been well studied in the univariate setting. Generally speaking, there are two different ways to construct a stopping rule. One is the CUSUM procedure (Page 1954; Lorden 1971) and  another is the Shiryayev-Roberts procedure (Shiryayev 1963; Roberts 1966). Let $f_0$ and $f_1$ be the probability density functions before and after a change point. The stopping rule for the CUSUM procedure is
\begin{eqnarray}
T_{{CUSUM}}=\inf \biggl\{n: \max_{0 \le k \le n} \sum_{i=k+1}^n \log \frac{f_1(X_i)}{f_0 (X_i)} > c_1 \biggr\}. \nonumber 
\end{eqnarray}  
The stopping rule for the Shiryayev-Roberts procedure is 
\begin{eqnarray}
T_{{SR}}=\inf \biggl\{n: \sum_{k=0}^{n-1} \sum_{i=k+1}^n \log \frac{f_1(X_i)}{f_0 (X_i)} > c_2 \biggr\}. \nonumber 
\end{eqnarray} 
The two procedures consider different ways to employ the likelihood ratio statistics. The CUSUM procedure terminates a monitored process if the maximum likelihood is more than a chosen threshold $c_1$. On the other hand, the Shiryayev-Roberts procedure weighs all the likelihood ratio statistics equally and terminates a process if the sum of them is more than a chosen threshold $c_2$. The optimality of the two procedures is investigated in different scenarios. The CUSUM procedure is minimax optimal (Lorden 1971; Moustakides 1986) and the Shiryayev-Roberts procedure is optimal when the change occurs at a distance time horizon ( Shiryaev 1961, 1963; Pollak and Tartakovsky 2009; Moustakides, Polunchenko and Tartakovsky 2011; Polunchenko and Tartakovsky 2010).    

The CUSUM and Shiryayev-Roberts procedures require both $f_0$ and $f_1$ are completely specified, which may not be realistic. To relax the restriction, Pollak and Siegmund (1991) considered a change-point detection in a normal mean when the initial mean is unknown but can be estimated from a training sample. 
If a normal mean after the change is unknown, Siegmund and Venkatraman (1995) proposed the generalized likelihood ratio statistic to detect a change point. As suggested by the authors, the proposed method can be combined with that in Pollak and Siegmund (1991) for a general setup of unknown initial mean and final mean. 

Due to the increasing availability of large scale data,  there has been growing interest in detecting abrupt
change in the multivariate and high-dimensional settings.
While some change point detection procedures were proposed to combine the CUSUM statistics from individual dimensions, they assumed the distributions before and after the change point are completely specified (Tartakovsky and Veeravalli 2008; Mei 2010). Some methods relaxed the restriction by only assuming the distribution after the change point to be unknown, but they simplified data dependence to independence across dimensions (Xie and Siegmund 2013). Except parametric methods, there also exist nonparametric procedures for sequential change point detection (Chen, 2019).     
In this article, we consider change point detection in the mean based on a more general setup that the distributions before and after the change point, and the covariance matrix across dimensions are all unknown. 
More precisely, we assume that $X_1, X_2, \cdots, X_{\tau}$ are independent $\mbox{N}_p(\mu_0, \Sigma)$ and $X_{\tau+1}, \cdots, $ are independent $\mbox{N}_p(\mu_1, \Sigma)$, where the dimension $p$ is large and $\tau$ is the change point if $\mu_0 \ne \mu_1$. Inspired by the classical CUSUM and Shiryayev-Roberts procedures, we propose two procedures to detect the change point $\tau$: one is based on a max-type U-statistic and another is based on a sum-type U-statistic. The proposed procedures assume the parameters $\mu_0$, $\mu_1$ and $\Sigma$ to be unknown, but there exists a training sample providing estimation of $\mu_0$ and $\Sigma$. The theoretical properties of the two procedures are explored in the high dimensional setting ($p \to \infty$). When there is no change point, we derive explicit expressions of the ARLs, which are employed to determine the thresholds of the two stopping rules with no need to run time-consuming Monte Carlo simulations. When there is a change point, we derive the EDDs as the functions of the magnitude of the change in the population mean, the threshold level and data dependence across components of the random vector $X$. The accuracy of the theoretical results is evaluated by simulation studies. 

The rest of the paper is organized as follows. Section 2 proposes the max-type and sum-type stopping rules and establish their relationship with the classical CUSUM and Shiryayev-Roberts procedures. Section 3 presents their asymptotic properties in the high dimensional setting. Simulation studies and real data analysis are given in Sections 4 and 5, respectively. We conclude the paper with some discussions in Section 6. The proofs of the main results are delegated to Appendix. 

\setcounter{section}{2} \setcounter{equation}{0}
\section*{\large 2. \bf Max-type and Sum-type Stopping Rules}

Let $X_1, \cdots, X_{\tau}$ be independent $\mbox{N}_p(\mu_0, \Sigma)$ and $X_{\tau+1}, \cdots, $ be independent $\mbox{N}_p(\mu_1, \Sigma)$, where 
$\tau$, $\mu_0$, $\mu_1$ and $\Sigma$ are all unknown.  
In the univariate setting ($p=1$), the problem of detecting $\tau$ was   
considered by Siegmund and Venkatraman (1995) where the authors assumed the univariate variance $\sigma^2=1$.  
More precisely, they proposed a statistic in the form  
\be
\frac{|t S_n/n -S_t |}{\{t(n-t)/n\}^{1/2}}, \quad \mbox{for} \,\, n_0 \le t <n, \label{LR}
\ee 
where $S_n=X_1+\cdots+X_n$, $S_t=X_1+\cdots+X_t$ and $n_0$ is the size of a training sample to provide an initial estimation of $\mu_0$. A detailed discussion about the derivation of (\ref{LR}) with a training sample can be seen in Pollak and Siegmund (1991). 

For the problem of detecting $\tau$ in the high-dimensional setting, we  consider the statistic 
\be
\frac{(t S_n/n -S_t)^{\prime}(t S_n/n -S_t)}{t(n-t)/n}, \quad \mbox{for} \,\, n_0 \le t <n, \label{LR-new}
\ee
which is the sum-of-squares of (\ref{LR}) from individual components.  It is worth noting that the above statistic is different from the sum-of-squares of a generalized likelihood ratio statistic in Xie and Siegmund (2013) where the authors assumed $\mu_0$ to be known. As what we mentioned above, when $\mu_0$ is unknown, a training sample is needed to provide an initial estimation of $\mu_0$ and leads to (\ref{LR-new}).   

A closer look at (\ref{LR-new}) shows that it involves the terms $X_i^{\prime} X_i$, which has the expectation $\mbox{tr}(\Sigma)$ and is irrelevant to the change-point detection in the mean. We therefore remove the terms $X_i^{\prime} X_i$ from (\ref{LR-new}) and propose a more efficient U-statistic  
\[
\frac{1}{n}\biggl(\frac{n-t}{t-1}\sum_{i \ne j=1}^{t} X_i^{\prime} X_j-2\sum_{i=1}^t \sum_{j=t+1}^n X_i^{\prime} X_j +\frac{t}{n-t-1} \sum_{i \ne j=t+1}^{n}X_i^{\prime} X_j\biggr).
\] 
A straightforward calculation shows that when the sample size $n$ is fixed, the expectation of the U-statistic equals zero when there is no change in the mean but positive when there is a change point. 

The above U-statistic consists of $X_n$ and its past observations $X_1, \cdots, X_{n-1}$. This may increase computational complexity especially with high-dimensional data. To improve the computational efficiency, we employ a moving window and propose a modified U-statistic
\begin{eqnarray}
    U_{t,n}(H) &=& \frac{1}{H} \biggl(\frac{n-t}{t-n+H-1}\sum_{i \ne j=n-H+1}^{t} X_i^{\prime} X_j-2\sum_{i=n-H+1}^t \sum_{j=t+1}^n X_i^{\prime} X_j \nonumber\\
    &+&\frac{t-n+H}{n-t-1} \sum_{i \ne j=t+1}^{n}X_i^{\prime} X_j\biggr), \label{stat1} 
\end{eqnarray}
which, compared with the original U-statistic, only employs the past $H$ observations $X_{n-H+1}, \cdots, X_n$ from the current $X_n$. The idea of using a moving window to reduce the computational complexity has been widely applied in the literature; see, for example, Lai (1995), and Cao et al. (2019).

With the U-statistic, we consider two methods to formulate a stopping rule. One is similar to the CUSUM which takes the maximum statistic. The obtained max-type stopping rule is 
\begin{eqnarray}
    T_1(H, a) = \text{inf}\Bigg\{n-n_0 : \max_{n-H+2 \le t \le n-2}\biggl|\frac{U_{t, n}(H)}{\tilde{\sigma}_{t}}\biggr |> a, n > n_0\Bigg\}. \label{rule1}
\end{eqnarray}
Another is similar to the Shiryayev-Roberts procedure which sums over all relevant statistics. The obtained sum-type stopping rule is
\begin{eqnarray}
    T_2(H, b) = \text{inf}\Bigg\{n-n_0 : \biggl | \sum_{t=n-H+2}^{n-2}\frac{U_{t,n}(H)}{\tilde{\sigma}} \biggr | > b, n > n_0\Bigg\}. \label{rule2}
\end{eqnarray}
When taking the maximum or the sum operations in the proposed stopping rules, we choose the range of $t$ between $n-H+2$ and $n-2$ rather than $n-H$ and $n$, because the U-statistic (\ref{stat1}) needs at least two observations to be formulated. Moreover, we standardize $U_{t,n}(H)$ and $\sum_{t=n-H+2}^{n-2}{U}_{t,n}(H)$ by $\tilde{\sigma}_{t}$ and $\tilde{\sigma}$, respectively, which estimate the standard deviations of the U-statistics for a fixed window size $H$. As shown in the following Proposition 1, we only need to estimate $\mbox{tr}(\Sigma^2)$, which can be achieved by employing a training sample of size $n_0$. 

\medskip
{\bf Proposition 1.} Assume $X_{1}, \cdots, X_H$ are independent $\mbox{N}_p(\mu_0, \Sigma)$, where $H$ is a fixed window length. The variance of $U_{t, n}(H)$ based on $\{X_1, \cdots, X_H\}$ is 
\begin{eqnarray}
\sigma_t^2&=&\biggl(\frac{H-t}{t-1}+2
+ \frac{t}{H-t-1} \biggr)\frac{2t(H-t)\,{\mbox{tr}({\Sigma}^2)}}{H^2}, \nonumber
\end{eqnarray}
and the variance of $\sum_{t=2}^{H-2}{U}_{t,n}(H)$ is
\begin{eqnarray}
\sigma^2=\sum_{t=2}^{H-2} \sigma_t^2. \nonumber
\end{eqnarray}

From Proposition 1, the only unknown in $\sigma_t^2$ and $\sigma^2$ is $\mbox{tr}(\Sigma^2)$. Based on the training sample $X_1, \cdots, X_{n_0}$, we estimate $\mbox{tr}(\Sigma^2)$ by 
\begin{eqnarray}
 \widetilde{\mbox{tr}({\Sigma}^2)}
&=& \frac{1}{n_0(n_0-1)} \sum_{i \ne j} (X_{i}^{\prime}
X_{j})^2-\frac{2}{n_0(n_0-1)(n_0-2)} \sum_{i,j,k}^{\star}
X_{i}^{\prime} X_{j}X_{j}^{\prime}
X_{k}\nonumber\\
&+&\frac{1}{n_0(n_0-1)(n_0-2)(n_0-3)} \sum_{i,j,k,l}^{\star}
X_{i}^{\prime} X_{j}X_{k}^{\prime} X_{l}, \label{var.est}
\end{eqnarray}
where $\sum^{\star}$ denotes summation over mutually distinct indices. The properties of (\ref{var.est}) especially its unbiasedness to $\mbox{tr}(\Sigma^2)$ can be seen in Li and Chen (2012). Replacing $\mbox{tr}(\Sigma^2)$ by (\ref{var.est}),  
we standardize $U_{t,n}(H)$ and $\sum_{t=n-H+2}^{n-2}{U}_{t,n}(H)$ by $\tilde{\sigma}_t$ and $\tilde{\sigma}$, respectively, to obtain the stopping rules (\ref{rule1}) and (\ref{rule2}) where 
\begin{eqnarray}
\tilde \sigma_t^2&=&\biggl(\frac{H-t}{t-1}+2
+ \frac{t}{H-t-1} \biggr) \frac{2t(H-t)\, \widetilde{\mbox{tr}({\Sigma}^2)}}{H^2}, \label{var.est1}
\end{eqnarray}
and 
\begin{eqnarray}
\tilde \sigma^2=\sum_{t=2}^{H-2}\tilde \sigma_t^2. \label{var.est2}
\end{eqnarray}

The proposed max-type or the sum-type stopping rules  terminate a monitored process if the corresponding statistic is greater than the threshold $a$ or $b$, respectively. In the sequential change point detection, the threshold should be chosen so that the ARL can be controlled at a pre-specified constant. 
In the next section, we derive explicit expressions of the ARL for both the max-type and sum-type stopping rules. Each expression relates the ARL with the corresponding threshold. As a result, the thresholds in the two stopping rules can be obtained by solving the equations rather than running time-consuming Monte Carlo simulations. When there is a change point, the performance of the max-type and sum-type stopping rules is evaluated by the EDD.   
In the next section, we derive their theoretical EDDs which are functions of the magnitude of the change in the population mean, the threshold level and data dependence across components of the random vector $X$. 


\section*{\large 3. \bf Asymptotic Results}

\subsection*{3.1 \bf Average run length (ARL)}

Let $\E_{\infty}$ and $\mbox{P}_{\infty}$ denote the expectation and probability, respectively, when there is no change. Let
\[
g(y, b)={2\log (y)}+1/2 \log \log (y)+\log (4/\sqrt{\pi})-b\sqrt{2\log (y)},
\]
and 
\[
s_{1y}= \frac{1}{y(1-y)}, \qquad s_{2y}=s_{1y}-2,  \qquad \nu(y)=\frac{(2/y)\{\Phi(y/2)-0.5\}}{(y/2) \Phi(y/2)+\phi(y/2)},
\]
where $\Phi(\cdot)$ and $\phi(\cdot)$ are the cumulative distribution function and the density function of the standard normal distribution, respectively. 

Note that the ARL is the expected value of a stopping time when there is no change. 
The following theorem establishes the ARLs for the proposed stopping rules (\ref{rule1}) and (\ref{rule2}).

\medskip
{\bf Theorem 1.} Assume that the eigenvalues of $\Sigma$ are bounded. As $p \to \infty$, and both $H$ and $a$ $\to \infty$ satisfying $H=O\{a^2\}$, 
\[
\mbox{E}_{\infty}\{T_1(H, a) \}=\frac{\sqrt{2\pi}\,\,H \,\,\mbox{exp}(a^2/2)}{ a^3 \int_{0}^1 s_{1y}\, s_{2y}\, \nu(a H^{-1/2} s_{1y}^{1/2}) \, \nu(a H^{-1/2} s_{2y}^{1/2})\,dy}\{1+o(1)\}.
\]
As $p \to \infty$, and both $H$ and $b$ $\to \infty$ satisfying $H=o\{\mbox{exp}(b^2/2)\}$, 
\[
\mbox{E}_{\infty}\{T_2(H, b) \}=\biggl( H+ \int_{H}^{\infty} \mbox{exp}\biggl[-\sqrt{2} \mbox{exp}\biggl\{g(t/H, b)\biggr\}\biggr] dt \biggr)\{1+o(1)\}.
\]

As shown in the proof of Theorem 1, the distribution of $T_1(H, a)$ is closely related with the maximum of the two-dimensional random field ${U_{t, n}(H)}/{\tilde{\sigma}_{t}}$. Similar to Siegmund and Venkatraman (1995), the expectation of $T_1(H, a)$ is obtained by establishing the asymptotical exponential distribution of $T_1(H, a)$, where the two functions $s_{1y}$ and $s_{2y}$ are the partial derivatives of the autovariance of the two-dimensional random field. On the other hand, the distribution of $T_2(H, b)$ is related with the maximum of the one-dimensional random field $\sum_{t=n-H+2}^{n-2}{U_{t,n}(H)}/{\tilde{\sigma}}$, which is shown to be the asymptotical Gumbel distribution. The expectation of $T_2(H, b)$ can be obtained accordingly.

Theorem 1 establishes the relationship between the ARL and the threshold for each stopping rule. The accuracy of the theoretical results is evaluated by simulation studies in the next section. To apply the max-type or the sum-type stopping rules to a change point detection process, we need to determine the value of the threshold. It can be obtained by solving the corresponding equation in Theorem 1 where the left hand side equals a pre-specified ARL.

\medskip
\subsection*{3.2 \bf Expected detection delay (EDD)}

When there is a change point, we evaluate the performance of the proposed stopping rules by the supremum conditional expected detection delay 
\[
\mbox{sup}_{n_0 \le \tau <\infty} \,\,\mbox{E}_{\tau}\{T_{1(2)}-(\tau-n_0)| T_{1(2)}>\tau-n_0\}.
\]
The supremum is achieved when the change occurs immediately after the training sample ($\tau=n_0$), and the corresponding expected detection delay is denoted by $\mbox{E}_0 \{T_{1(2)}\}$. The following theorem establishes the asymptotic results of $\mbox{E}_0 \{T_{1(2)}\}$. 

\medskip
{\bf Theorem 2.} Let $\Delta^2=(\mu_1-\mu_0)^{\prime}(\mu_1-\mu_0)$. As $p$, $H$ and $a$ $\to \infty$, 
\[
\frac{a {\sigma}_{\min}+\rho_{\min}(\Delta)}{\Delta^2}\le\mbox{E}_{0}\{T_1(H, a) \}\le \frac{a {\sigma}_{\max}+\rho_{\max}(\Delta)}{\Delta^2},
\]
where ${\sigma}_{\min}=\min_{n-H+2 \le t \le n-2} {\sigma}_t$ and ${\sigma}_{\max}=\max_{n-H+2 \le t \le n-2} {\sigma}_t$ with $\sigma_t$ given in Proposition 1, and $\rho_{\min}(\Delta)$ and $\rho_{\max}(\Delta)$ are the expected overshoots of the stopping rule (\ref{rule1}) over the boundary $a$ given by $$\rho_{\min}(\Delta)=\mbox{E}_0\{ \max_{n-H+2 \le t \le n-2}|{U_{t,n}(H)} |- a\,\, {{\sigma}_{\min}}\},$$ $$\rho_{\max}(\Delta)=\mbox{E}_0\{ \max_{n-H+2 \le t \le n-2}|{U_{t,n}(H)} |- a\,\, {{\sigma}_{\max}}\}.$$

As $p$, $H$ and $b$ $\to \infty$, 
\[
\mbox{E}_{0}\{T_2(H, b) \}=\frac{b {\sigma}+\rho_2(\Delta)}{H \Delta^2},
\]
where the expected overshoot of the stopping rule (\ref{rule2}) over the boundary is $$\rho_2(\Delta)=\mbox{E}_0\{  | \sum_{t=n-H+2}^{n-2}{U_{t,n}(H)} | - b \, {{\sigma}} \}.$$

Theorem 2 demonstrates the impact of the magnitude of the change in the population mean, the threshold level and data dependence on the EDDs of the proposed max-type and sum-type procedures. If $\mu_0$ is known and $\Sigma=I_p$, the EDD of the proposed max-type stopping rule can be reduced to the leading order result of Xie and Siegmund (2013). Our result is more general as it is based on the assumption that $\mu_0$, $\mu_1$ and $\Sigma$ are all unknown. Moreover, we derive the EDD of the sum-type procedure so that the performance of the two procedures can be compared when there is a change point.  
   
In the univariate setting, the numerical comparison in Moustakides, Polunchenko and Tartakovsky (2009) has shown that the CUSUM procedure outperforms the Shiryayev-Roberts procedure in terms of the supremum conditional expected detection delay. 
Since we do not have explicit expressions for the expected overshoots $\rho_{\min}(\Delta)$, $\rho_{\max}(\Delta)$ and $\rho_{2}(\Delta)$, we compare the proposed procedures numerically in the next section, and show that 
there is a preference for the proposed max-type procedure in terms of the supremum conditional expected detection delay. Note that the proposed max-type and sum-type procedures can be thought as the analog of the CUSUM and Shiryayev-Roberts procedures. Our results therefore extend the finding in Moustakides, Polunchenko and Tartakovsky (2009) from the univariate setting to the high dimensional setting.  
\section*{\large 4. \bf Simulation Studies}

\subsection*{4.1 \bf Accuracy of ARLs}

We first evaluate the performance of the proposed max-type and sum-type procedures when there is no change in the population mean. 
Each random vector $X_{i}$ is generated from $\mbox{N}_p(\mu, \Sigma)$.
We choose $\mu=0$ for simplicity, and consider the covariance matrix $\Sigma_{ij}=0.5^{|i-j|}$ for $1\le i, j\le p$.  
We choose the dimension $p=1000$, $1500$, $2000$ and $2500$. Note that the proposed detection procedures need a training sample and window size. Here we choose the training sample size $n_0=200$, and the window-size $H=100$.

To examine the accuracy of Theorem 1, we choose nominal ARLs to be $1000$, $3000$, $5000$ and $7000$, respectively. For each nominal ARL, we solve the two equations in Theorem 1 to obtain the thresholds $a$ and $b$, respectively.  Based on the obtained $a$ or $b$, we obtain the Monte Carlo ARLs by averaging the stopping times of the max-type or sum-type procedure over $1000$ simulations.    
Table \ref{case1} demonstrates the Monte Carlo ARLs corresponding to nominal ARLs. As we can see, all the Monte Carlo ARLs are reasonably close to the nominal ARLs, confirming the accuracy of Theorem 1. 

\begin{table}[t!]
\tabcolsep 8pt
\begin{center}
\caption{The accuracy of derived theoretical ARLs for both Max-type and Sum-type procedures. The Monte Carlo ARLs are obtained based on $1000$ simulations. }
\label{case1}
\begin{tabular}{cccccccccccc}
\\
\hline  &\multicolumn{2}{c}{$p=1000$}&&\multicolumn{2}{c}{$p=1500$} &&\multicolumn{2}{c}{$p=2000$} &&\multicolumn{2}{c}{$p=2500$}\\[1mm]
\cline{2-3} \cline{5-6} \cline{8-9} \cline{11-12}  ARL & Max & Sum && Max & Sum &  & Max & Sum && Max & Sum \\\hline
$1000$     & $1074$ & $914$ & & $1151$ & $966$ & & $1210$ & $932$&& $1290$ & $906$\\
$3000$     & $2561$ & $2570$ & & $2884$ & $2805$ & & $3012$ & $2921$&& $3276$ & $2870$\\
$5000$     & $3517$ & $4572$ & & $4428$ & $4771$ & & $4757$ & $4812$&& $4934$ & $5198$\\
$7000$     & $4781$ & $6152$ & & $5913$ & $6871$ & & $6480$ & $6611$&& $6867$ & $7254$\\
\hline
\end{tabular}
\end{center}
\end{table}

\subsection*{4.1 \bf Comparison of EDDs}


We then compare the max-type and sum-type procedures when there is a change point. The training sample size is $200$ and window-size is $100$. The change point is chosen to occur immediately after the training sample ($\tau=200$). 
For the sake of simplicity, the population mean before the change point is zero. The population mean after the change point is a nonzero vector $\mu$ with dimension $p=2000$. We let $\delta=(\mu^{\prime}\mu)^{1/2}$ and choose $\delta=5, 10, 15, 20$ representing different magnitudes of change in population mean.  

\begin{figure}[t!]
\begin{center}
\includegraphics[width=0.45\textwidth,height=0.43\textwidth]{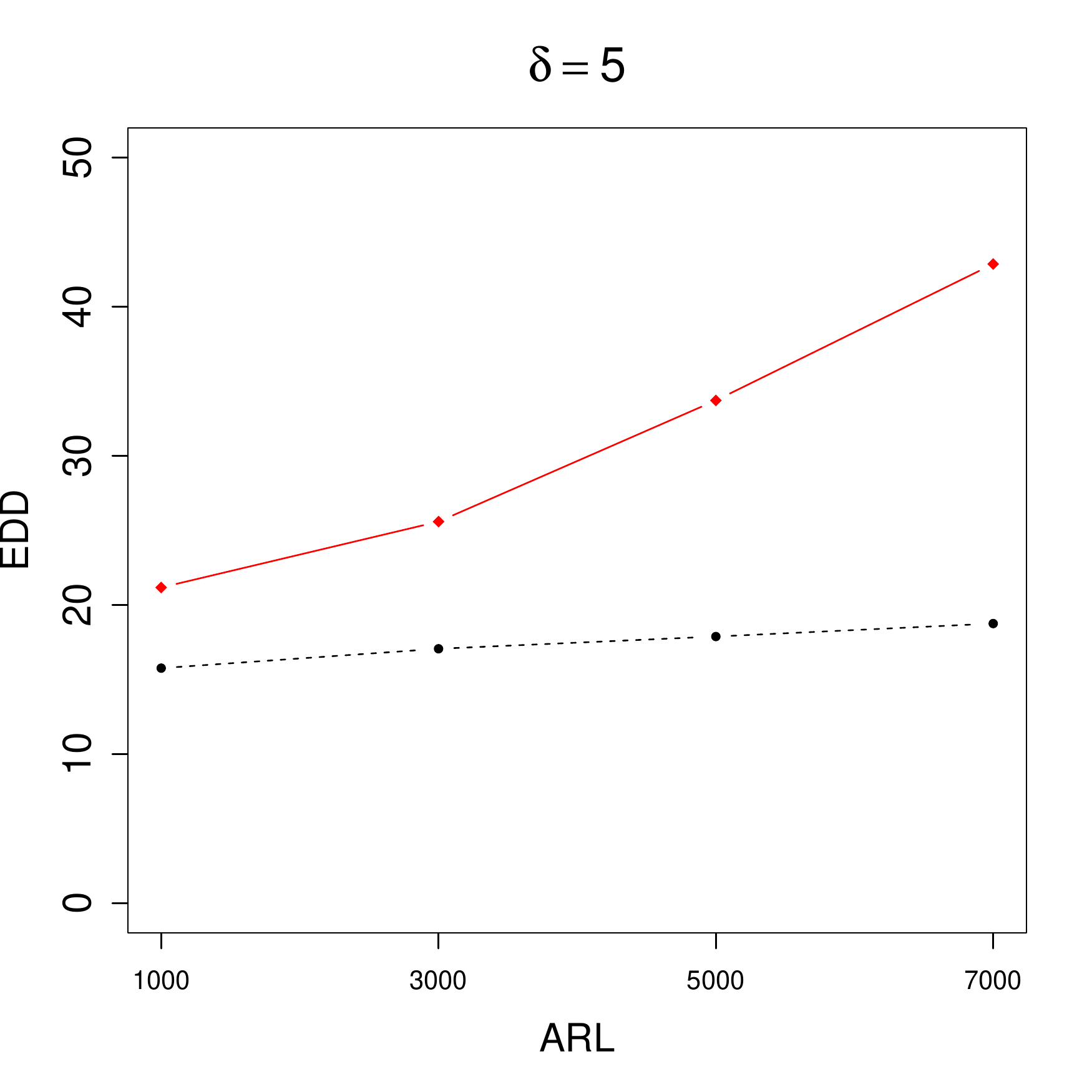}
\includegraphics[width=0.45\textwidth,height=0.43\textwidth]{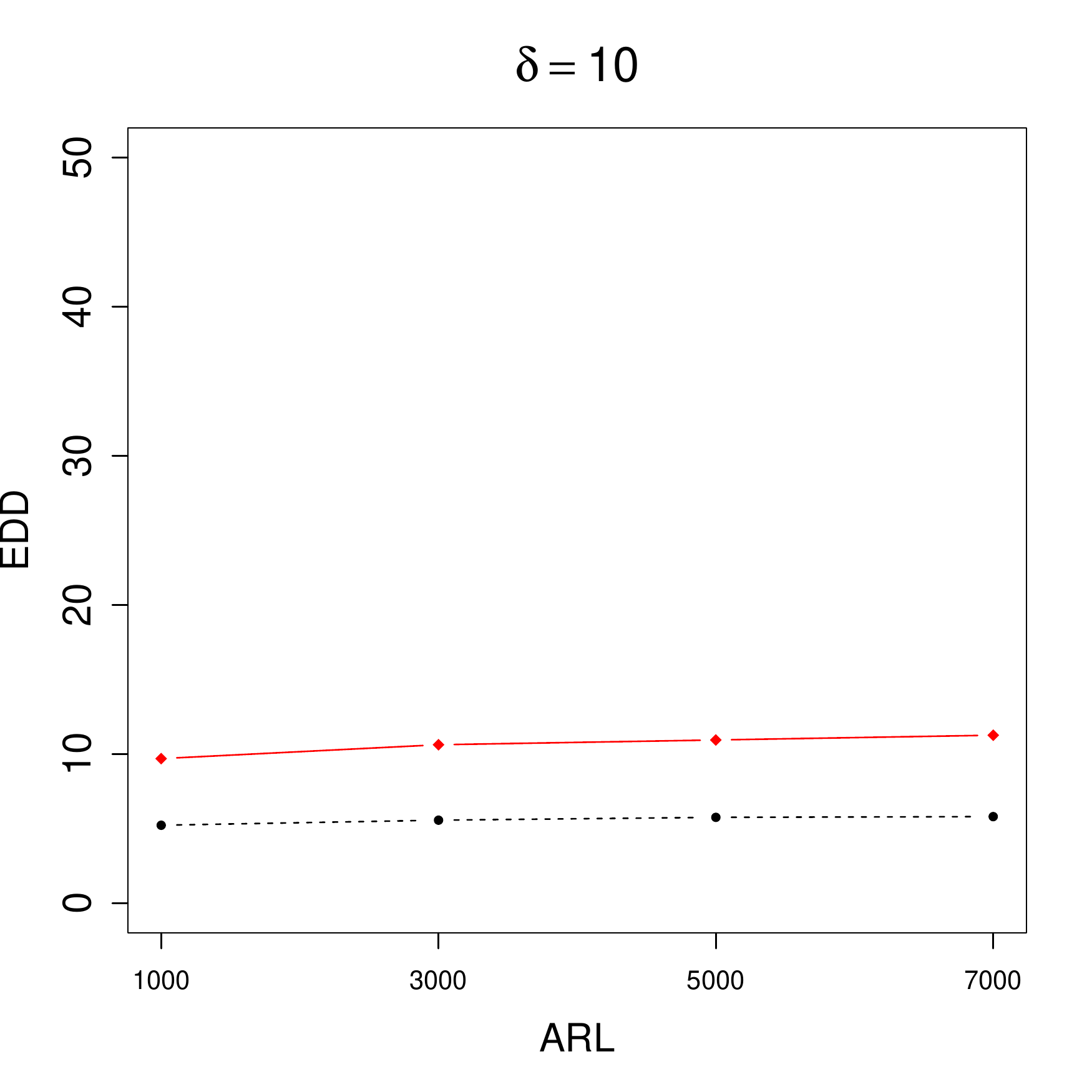}\\
\includegraphics[width=0.45\textwidth,height=0.43\textwidth]{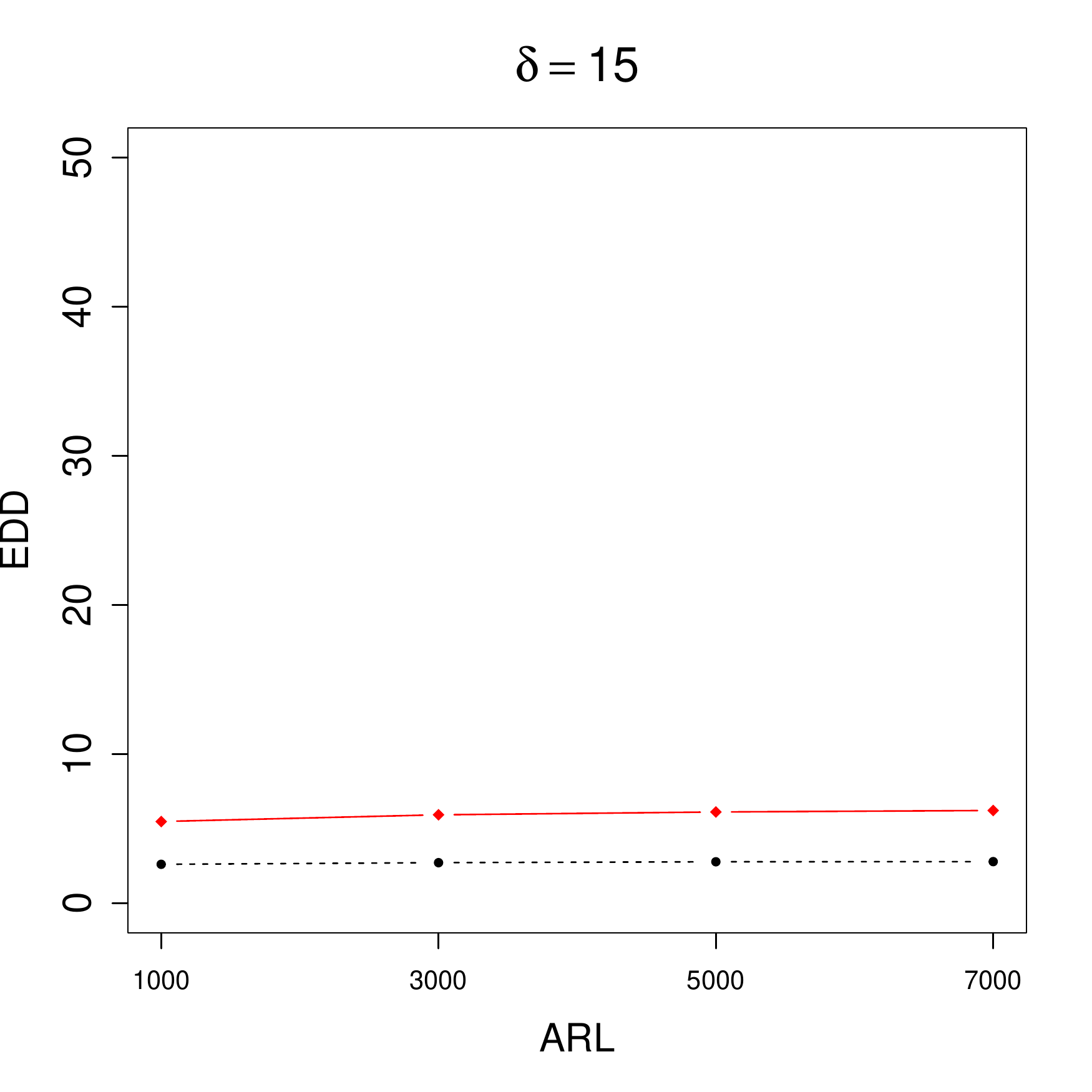}
\includegraphics[width=0.45\textwidth,height=0.43\textwidth]{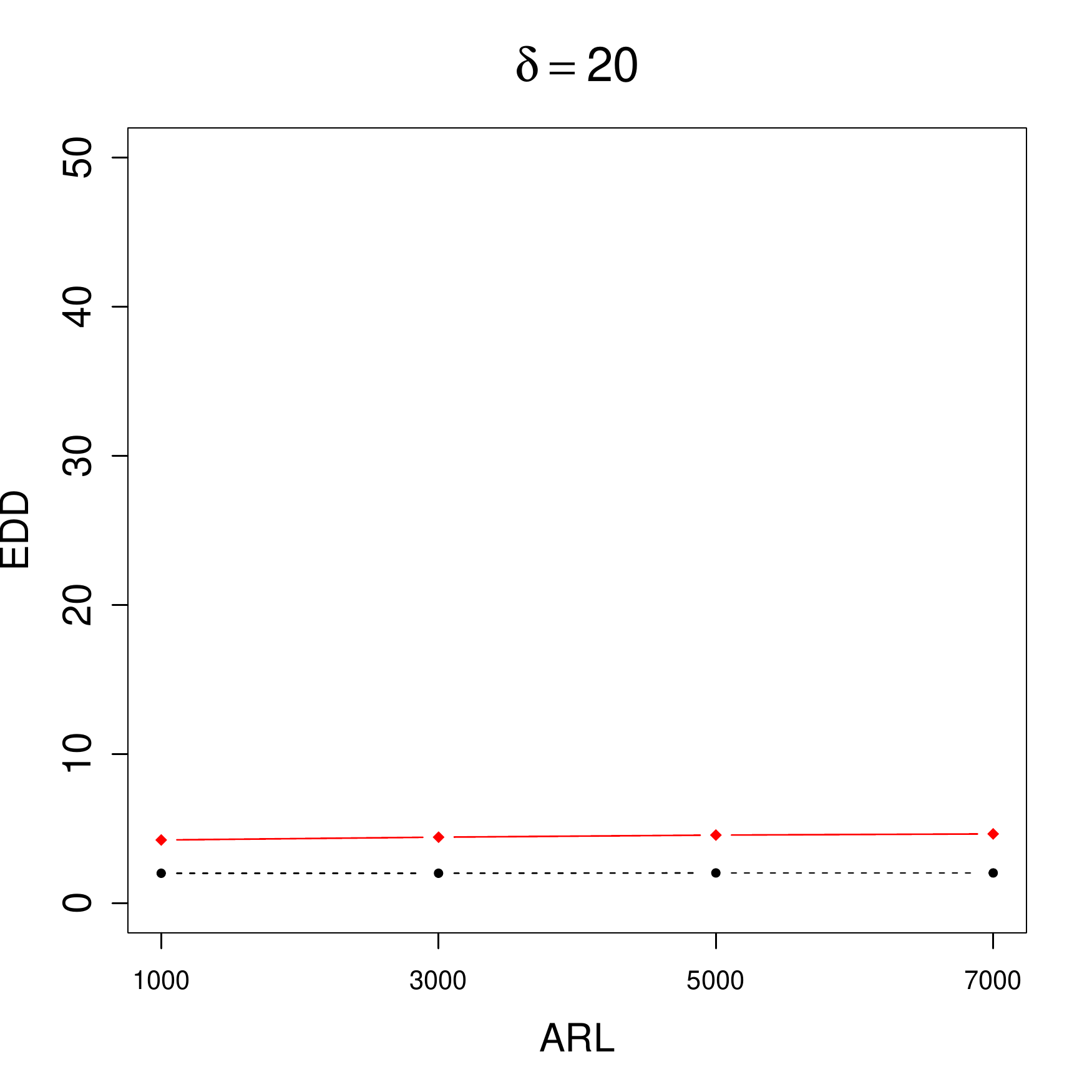}
\caption{The EDDs of max-type and sum-type procedures with different change magnitudes $\delta$. In each panel, the dashed line and the solid line represent the max-type procedure and the sum-type procedure, respectively. }
\label{fig1}
\end{center}
\end{figure}

It is fair to compare the two procedures with a change point if both of them control the ARL at the same number. 
We consider the nominal ARLs to be $1000$, $3000$, $5000$ and $7000$, respectively.  For each nominal ARL, we obtain the thresholds of the max-type and sum-type procedures, respectively, by solving the two equations in Theorem 1. 
With the obtained thresholds, the EDDs of the max-type and sum-type procedures are obtained by averaging the stopping times of each procedure from $1000$ simulations. 
Figure \ref{fig1} compares the EDDs of the max-type (dashed line) and sum-type procedures (solid line) for different magnitudes $\delta$. 
As the magnitude of change $\delta$ increases, the EDDs of the max-type and sum-type procedures decrease. 
With smaller EDDs, the max-type procedure outperforms the sum-type procedure especially when $\delta$ is smaller. As $\delta$ becomes greater, the two procedures perform closer to each other. 

\section*{\large 5. \bf Application to real data}

\begin{figure}[t!]
\begin{center}
\includegraphics[width=0.49\textwidth,height=0.43\textwidth]{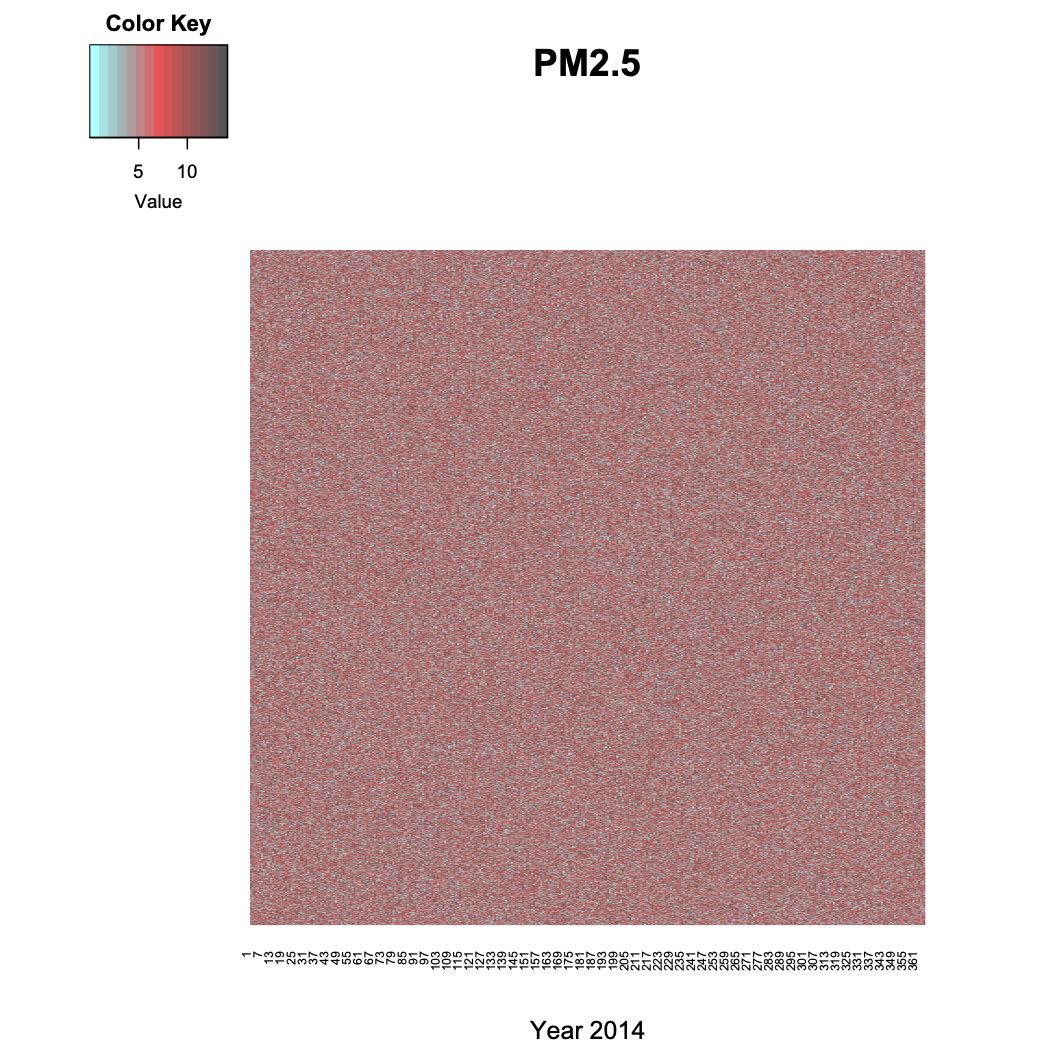}
\includegraphics[width=0.49\textwidth,height=0.43\textwidth]{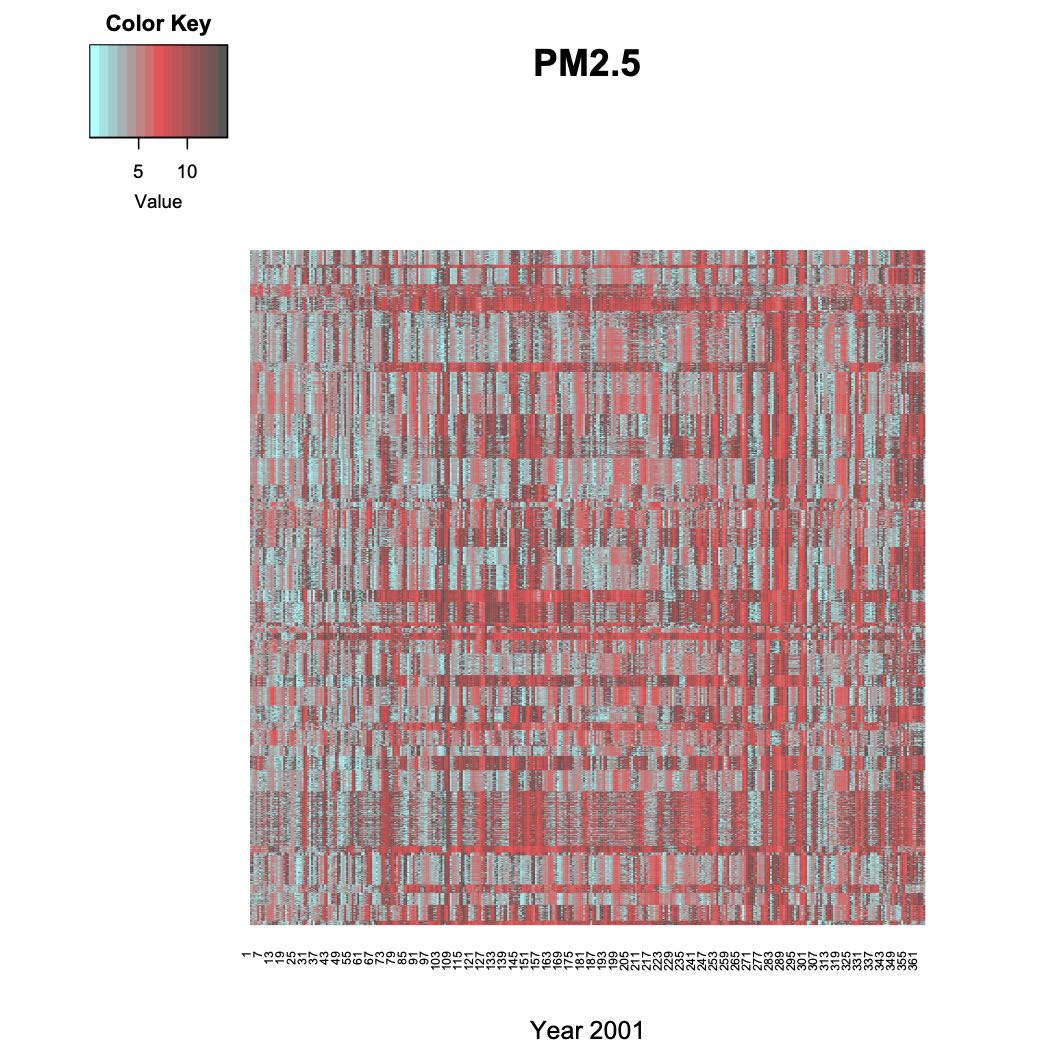}
\caption{Heatmap of PM2.5 concentrations at 3109 sites for the United States. Left panel: PM2.5 concentrations in 2014; Right panel: PM2.5 concentrations in 2001.}
\label{fig2}
\end{center}
\end{figure}

To demonstrate the practical use of the max-type and sum-type procedures, 
we consider a daily PM2.5 concentration dataset which is accessible at Centers for Disease Control and Prevention, National Environmental Public Health Tracking Network. PM2.5 poses a significant risk to human health because its microscopic diameter makes it easy to be absorbed into the bloodstream. The data provide hourly PM 2.5 concentration at 3109 sites ($p=3109$) for the continental United States, from January 1, 2001 to December 31, 2014.  

According to United States Environmental Protection Agency, average PM2.5 concentrations in the continental United States have decreased 43\%  since 2000. To illustrate this trend, we compare the heatmap of the PM2.5 concentrations in 2014 with that in 2001. As we can see in Figure \ref{fig2}, there is a homogeneity in PM2.5 concentrations across 3109 sites over 2014 but a strong heterogeneity over 2001. To exam if the proposed max-type and sum-type procedures are able to identify such a change, we artificially combine the PM2.5 concentrations in 2014 with those in 2001, and employ the first 100 observations in 2014 as the training data because they show the homogeneity in the mean. We choose the thresholds $a$ and $b$ of the two stopping rules to be $4.60$ and $3.58$ respectively, so that the nominal ARL is controlled at $7000$. Both the max-type and sum-type procedures run through year 2014 without terminating the process, but detect a change at January 2nd and 3rd of 2001, respectively. The identified change points represent the transitioning of PM2.5 concentrations from homogeneity to heterogeneity.       

\section*{\large 6. \bf Conclusion and discussion}

For the problem of detecting a change point in the mean, we propose a max-type procedure and a sum-type procedure. The proposed procedures assume the population mean before and after the change point, and the covariance matrix across components of a random vector are all unknown. To study theoretical properties of the proposed procedures, we derive their average run lengths (ARLs) and expected detection delays (EDDs). The numerical studies confirm the accuracy of the derived ARLs, and suggest a preference for the proposed max-type procedure.  

While we assume the random vectors are normally distributed, the proposed procedures are expected to work for non-Gaussian data as well. Similar to Li and Li (2019), more technical conditions are needed to establish the theoretical results when data are non normally distributed. The proposed max-type and sum-type procedures in the high dimensional setting are the analog of the CUSUM and Shiryayev-Roberts procedures, respectively, in the univariate setting. The previous numerical studies show that the CUSUM procedure outperforms the Shiryayev-Roberts procedure in terms of the supremum conditional expected detection delay. Our numerical investigation achieves the same conclusion for the proposed max-type and sum-type procedures. 

\section*{\large Appendix: Technical Details.}

\bigskip
\noindent{\bf A.1. Proof of Proposition 1.}
\bigskip

Based on $\{X_1, \cdots, X_H \}$, we have
\begin{eqnarray}
    U_{t,n}(H) &=& \frac{1}{H} \biggl(\frac{H-t}{t-1}\sum_{i \ne j=1}^{t} X_i^{\prime} X_j-2\sum_{i=1}^t \sum_{j=t+1}^H X_i^{\prime} X_j +\frac{t}{H-t-1} \sum_{i \ne j=t+1}^{H}X_i^{\prime} X_j\biggr). \nonumber
\end{eqnarray}
The variance of $U_{t,n}(H)$ is
\[
\sigma_t^2=\mbox{E}\{U_{t,n}^2(H) \}-\mbox{E}^2\{U_{t,n}(H)\}.
\]
A direct computation can show that 
\[
\mbox{E}\{U_{t,n}(H)\}=0. 
\]

Since $U_{t,n}(H)$ is scale invariant, we can simply assume $\mbox{E}(X_i)=0$. Moreover, 
\begin{eqnarray}
    U_{t,n}^2(H) &=& \frac{1}{H^2} \biggl\{\frac{(H-t)^2}{(t-1)^2}\sum_{i \ne j=1}^{t} \sum_{k \ne l=1}^{t} X_i^{\prime} X_j X_k^{\prime} X_l+4\sum_{i=1}^t \sum_{j=t+1}^H \sum_{k=1}^t \sum_{l=t+1}^H X_i^{\prime} X_j X_k^{\prime} X_l\nonumber\\
    &+&\frac{t^2}{(H-t-1)^2} \sum_{i \ne j=t+1}^{H} \sum_{k \ne l=t+1}^{H}X_i^{\prime} X_j X_k^{\prime} X_l\biggr\}. \nonumber
\end{eqnarray}
Using the result that $\mbox{E}(X_1^{\prime} X_2 X_2^{\prime} X_1)=\mbox{tr}(\Sigma^2)$, we can derive the expression of $\sigma_t^2$. 
This completes the proof of Proposition 1.

\bigskip
\noindent{\bf A.2. Proof of Theorem 1.}
\bigskip

We first derive the ARL of the max-type stopping rule. From (\ref{rule1}), the cumulative distribution function of $T_1(H, a)$ is
$$\mbox{P}_{\infty}\{T_1(H, a) \le m \}=\mbox{P}_{\infty}\biggl\{\max_{n \le m, n-H+2 \le t \le n-2}\biggl|\frac{U_{t,n}(H)}{\tilde{\sigma}_{t}}\biggr |> a \biggr\}.$$  
To obtain the distribution of $T_1(H, a)$, we first derive the distribution of ${U_{t,n}(H)}/{\tilde{\sigma}_{t}}$ for a given $t$ and $n$. We assume that data are normally distributed and the eigenvalues of the covariance matrix $\Sigma$ are bounded. The assumptions are sufficient to the conditions in Chen and Qin (2010). As a result, ${U_{t,n}(H)}/{\tilde{\sigma}_{t}}$ follows the standard normal according to Theorem 1 of Chen and Qin (2010).   

Next from arguments in Theorem 1 of Siegmund and Venkatraman (1995), Theorem 1 of Xie and Siegmund (2013) and Theorem 4.2 of Chen (2019), ${U_{t,n}(H)}/{\tilde{\sigma}_{t}}$ is the two-dimensional random field for $t \le n-2$, and when both $H$ and $a \to \infty$ satisfying $H=O\{a^2\}$, $T_1(H, a)$ is asymptotically exponentially distributed with parameter $\lambda$ equal to 
$$\phi(a) (a/\sqrt{H})^2 \,a \int_0^1 s_{1y} s_{2y} \nu(a/\sqrt{H} \sqrt{s_{1y}}) \nu(a/\sqrt{H} \sqrt{s_{2y}})\,dy,$$
where $\phi(x)$ is the standard normal density function, $\Phi(x)$ is the cumulative distribution function, and $\nu(x)=2x^{-2} \mbox{exp}\{-2 \sum_{n=1}^{\infty} n^{-1} \Phi(-x/2 n^{1/2}) \}$. Moreover, letting 
$$\rho(\epsilon_1, \epsilon_2)=\mbox{cov}\biggl(\frac{U_{n-yH,n}}{\tilde{\sigma}_{n-yH}}, \,\,\frac{U_{n-yH+\epsilon_1H,n+\epsilon_2H}}{\tilde{\sigma}_{n-yH+\epsilon_1H}}\biggr),$$
we have 
$$s_{1y}=-\frac{\partial \rho(\epsilon_1, 0)}{\partial \epsilon_1} \biggl|_{\epsilon_1=0}, \quad s_{2y}=-\frac{\partial \rho(0,\epsilon_2)}{\partial \epsilon_2} \biggl|_{\epsilon_2=0}.$$

Using (\ref{stat1}), we can derive
\begin{eqnarray}
\rho(\epsilon_1, \epsilon_2)&=& \frac{y(1-y-\epsilon_2)^2(\epsilon_2-\epsilon_1+y)}{(1-y)(1-\epsilon_2+\epsilon_1-y)}+\frac{\epsilon_1^2(1-y)(\epsilon_2-\epsilon_1+y)}{y(1-\epsilon_2+\epsilon_1-y)}-\frac{2\epsilon_1(1-y)(y-\epsilon_1)}{y}\nonumber\\
&+&\frac{(y-\epsilon_1)^2(1-y)(1+\epsilon_1-\epsilon_2-y)}{y(\epsilon_2-\epsilon_1+y)}-\frac{2\epsilon_1(\epsilon_2-\epsilon_1+y)(1-y-\epsilon_2)}{1-\epsilon_2+\epsilon_1-y}\nonumber\\
&+&2(1-y-\epsilon_2)(y-\epsilon_1).\nonumber
\end{eqnarray}
Then, 
$$s_{1y}=\frac{1}{y(1-y)}, \quad s_{2y}=\frac{1}{y(1-y)}-2.$$
Using the result for the expectation of exponential distribution, we can derive the ARL of the max-type stopping rule, which is the first result of Theorem 1.

We then derive the ARL of the sum-type stopping rule. From (\ref{rule2}), the cumulative distribution function of $T_2(H, b)$ is
$$\mbox{P}_{\infty}\{T_2(H, b) \le m \}=\mbox{P}_{\infty}\biggl\{\max_{0 \le n \le m}\biggl|\sum_{t=n-H+2}^{n-2}\frac{U_{t,n}(H)}{\tilde{\sigma}}\biggr |> b \biggr\}.$$ 
Similar to the proof for the max-type stopping rule,  $\sum_{t=n-H+2}^{n-2}{U_{t,n}(H)}/{\tilde{\sigma}}$ follows the standard normal when data are normally distributed and the eigenvalues of the covariance matrix $\Sigma$ are bounded. 
Then from the arguments for Theorem 1 of Li and Li (2020), $\max_{0 \le n \le m}\biggl|\sum_{t=n-H+2}^{n-2}{U_{t,n}(H)}/{\tilde{\sigma}}\biggr |$ is asymptotically Gumbel distributed when both $H$ and $b \to \infty$ satisfying $H=o\{\mbox{exp}(b^2/2)\}$. As a result, the cumulative distribution function
$$\mbox{P}_{\infty}\{T_2(H, b) \le m \}= 1- \mbox{exp}\biggl[-\sqrt{2}\mbox{exp}\biggl\{g(m/L,b)\biggr\} \biggr].$$
The ARL of the sum-type stopping rule can be obtained accordingly, which is the second result of Theorem 1. 

\bigskip
\noindent{\bf A.3. Proof of Theorem 2.}
\bigskip

We first derive the EDD of the max-type stopping rule (\ref{rule1}). Toward this end, we notice that 
\[
\mbox{E}_0\{T_{\min}(H,a)\} \le \mbox{E}_0\{T_{1}(H,a)\} \le \mbox{E}_0\{T_{\max}(H,a)\},  
\]
where  
$$T_{\min}(H, a) = \text{inf}\Bigg\{n-n_0 : \max_{n-H+2 \le t \le n-2}|{U_{t,n}(H)} |> a {\tilde{\sigma}_{\min}}, n > n_0\Bigg\},$$
and
$$T_{\max}(H, a) = \text{inf}\Bigg\{n-n_0 : \max_{n-H+2 \le t \le n-2}|{U_{t,n}(H)} |> a {\tilde{\sigma}_{\max}}, n > n_0\Bigg\}.$$

To find $\mbox{E}_0\{T_{\min}(H,a)\}$, we need to evaluate $\mbox{E}_0 \biggl\{\max_{n-H+2 \le t \le n-2}|{U_{t,n}(H)} \biggr\}|$ whose leading order is the same as 
$$\max_{n-H+2 \le t \le n-2}\frac{H}{(t-n+H)(n-t)}\biggl(\frac{t-n+H}{H}S_n-S_t\biggr)^{\prime}\biggl(\frac{t-n+H}{H}S_n-S_t\biggr),$$ 
which can be also written as the expectation of 
\begin{eqnarray}
&&\max_{n-H+2 \le t \le n-2}\biggl[\frac{2(H-n+n_0)}{H} \Delta\mu^{\prime}(S_n-\frac{n-n_0}{2}\Delta\mu)-\frac{2(H-n+n_0)}{t-n+H} \Delta\mu^{\prime}(S_t-\frac{t-n_0}{2}\Delta\mu)\nonumber\\
&+&\frac{H}{(t-n+H)(n-t)}\biggl\{\frac{t-n+H}{H}S_n-S_t-\frac{(n-t)(H-n+n_0)}{H}\Delta\mu\biggr\}^{\prime}\nonumber\\
&&\biggl\{\frac{t-n+H}{H}S_n-S_t-\frac{(n-t)(H-n+n_0)}{H}\Delta\mu\biggr\}\biggr],\nonumber
\end{eqnarray}
where $\Delta\mu=\mu_1-\mu_0$. 

Similar to Siegmund (1985) and Xie and Siegmund (2013), the second term above can be shown to follow a random walk and the third term is asymptotically normally distributed whose expectation does not depend on $\Delta \mu$. The leading order is contributed by the first term ${2(H-n+n_0)}{H^{-1}} \Delta\mu^{\prime}(S_n-\frac{n-n_0}{2}\Delta\mu)$. 
Using the Wald's identity and  as $H \to \infty$, 
$$\mbox{E}_0\{{2(H-n+n_0)}{H^{-1}}\Delta\mu^{\prime}(S_n-\frac{n-n_0}{2}\Delta\mu)\}= \Delta^2 \mbox{E}_0\{T_{\min}(H,a)\}\{1+o(1)\},$$
where $T_{\min}=n-n_0$ and $\Delta^2=\Delta \mu^{\prime} \Delta \mu$.

Next, we write
$$\mbox{E}_0\biggl(\max_{n-H+2 \le t \le n-2}|{U_{t,n}(H)} |\biggr)=a \sigma_{\min}+\mbox{E}_0\biggl(\max_{n-H+2 \le t \le n-2}|{U_{t,n}(H)} |-a \tilde{\sigma}_{\min}\biggr).$$
Applying the nonlinear renewal theory in Siegmund (1985), we let $\rho_{\min}(\Delta)$ be the expected overshoot of the stopping rule over the boundary, which is
$$\mbox{E}_0\biggl(\max_{n-H+2 \le t \le n-2}|{U_{t,n}(H)} |-a \tilde{\sigma}_{\min}\biggr)=\rho_{\min}(\Delta).$$

Using the established equation that
$$a \sigma_{\min}+\rho_{\min}(\Delta)= \Delta^2 \mbox{E}_0\{T_{\min}(H,a)\}\{1+o(1)\},$$
we have
$$\mbox{E}_0\{T_{\min}(H,a)\}=\frac{a \sigma_{\min}+\rho_{\min}(\Delta)}{\Delta^2}\{1+o(1)\}.$$
Using the same approach, we can derive
$$\mbox{E}_0\{T_{\max}(H,a)\}=\frac{a \sigma_{\max}+\rho_{\max}(\Delta)}{\Delta^2}\{1+o(1)\}.$$
This completes the first part of Theorem 2. 

After changing the operation of maximum into the sum and repeating the above procedure, we can derive the EDD of the sum-type procedure. This completes the second part of Theorem 2. 

\section*{Reference}

\begin{description}
\setlength{\baselineskip}%
{0.6\baselineskip}
\setlength{\itemsep}{1.5pt}%

\item
{Cao, Y., Thompson, A., Wang, M. and Xie, Y.}. ``Sketching for sequential change-point detection,''
\textit{EURASIP Journal on Advances in Signal Processing}, 42.  

\item
Chen, H. (2019), ``Sequential change-point detection based on nearest neighbors," \textit{The Annals of Statistics}, 47, 1381-1407.

\item
Chen, S. X. and Qin, Y. L. (2010), ``A two-sample test for high-dimensional data with applications to gene-set testing," \textit{The Annals of Statistics}, 38, 808-835.

\item
Lai, T. (1995), ``Sequential changepoint detection in quality control and dynamical systems," \textit{Journal of Royal Statistical Society, Series B}, 57, 613–658.

\item
Li, J. and Chen, S. X. (2012), ``Two sample tests for high-dimensional covariance matrices," \textit{The Annals of Statistics}, 40(2), 908-940.

\item
Li, L. and Li, J. (2019), ``Change point detection in the mean of high-dimensional covariance structure with application to dynamic networks," arXiv:1911.07762.

\item
Lorden, G. (1971), ``Procedures for reacting to a change in distribution," \textit{The Annals of Mathematical Statistics}, 42(6), 1897-1908.

\item
Mei, Y. (2010), ``Efficient scalable schemes for monitoring a large number of data streams," \textit{Biometrika}, 97(2), 419-433.

\item 
Moustakides, G. V., Polunchenko, A. S. and Tartakovsky, A. G. (2009). ``Numerical comparison of CUSUM and Shiryaev-Roberts procedures for detecting changes in distributions,''  \textit{Communications in Statistics - Theory and Methods}, 38, 3225-3239.

\item 
Moustakides, G. V., Polunchenko, A. S. and Tartakovsky, A. G. (2011). ``A numerical approach to performance analysis of change-point detection procedures,''  \textit{Statistica Sinica}, 21, 571-596.

\item
Page, E. S. (1954), ``Continuous Inspection Schemes," \textit{Biometrika}, 41(1/2), 100-115.

\item 
Pollak, M. and Siegmund, D. (1991). ``Sequential detection of a change in a normal mean when the initial value is unknown,''  \textit{The Annals of Statistics}, 19, 394-416.

\item 
Pollak, M. and Tartakovsky, A. G. (2009). ``Optimality properties of the Shiryaev-Roberts procedure,''  \textit{Statistica Sinica}, 19, 1729-1739.

\item 
Polunchenko, A. S. and Tartakovsky, A. G. (2010). ``On optimality of the Shiryaev-Roberts procedure for detecting a change in distribution,''  \textit{The Annals of Statistics}, 38, 3445-3457.

\item
Roberts, S. W. (1966), ``A comparison of some control chart procedures," \textit{Technometrics}, 8(3), 411-430.

\item
Shiryayev, A. N. (1961), ``The problem of the most rapid detection of a disturbance in a stationary process," \textit{Soviet Math. Dokl.}, 2, 795-799.

\item
Shiryayev, A. N. (1963), ``On optimal methods in earliest detection problems," \textit{Theory of Probability and its Applications}, 8, 26-51.

\item
Siegmund, D. (1985), \textit{Sequential analysis: tests and confidence intervals}, Springer Science \& Business Media.

\item
Siegmund, D. and Venkatraman, E. S. (1995), ``Using the generalized likelihood ratio statistic for sequential detection of a change-point," \textit{The Annals of Statistics}, 255-271.

\item
Tartakovsky, A. G. and Veeravalli, V. V. (2008), ``Asymptotically optimal quickest change detection in distributed sensor systems," \textit{Sequential Analysis}, 27(4), 441-475.
\item
Xie, Y. and Siegmund, D. (2013), ``Sequential multi-sensor change-point detection," \textit{Annals of Statistics}, 41, 670-692.

\end{description}

\end{document}